\documentclass[pdflatex,sn-mathphys-num,iicol]{sn-jnl}

\usepackage{graphicx}%
\usepackage{multirow}%
\usepackage{amsmath,amssymb,amsfonts}%
\usepackage{amsthm}%
\usepackage{mathrsfs}%
\usepackage[title]{appendix}%
\usepackage{xcolor}%
\usepackage{textcomp}%
\usepackage{manyfoot}%
\usepackage{booktabs}%
\usepackage{algorithm}%
\usepackage{algorithmicx}%
\usepackage{algpseudocode}%
\usepackage{listings}%
\usepackage{todonotes}
\usepackage{textcomp}
\usepackage[utf8]{inputenc}
\usepackage{xfrac}
\usepackage{subcaption}
\usepackage[switch]{lineno}
\begin{document}

\title[Measurements of the Quantum Yield of Silicon using Geiger-mode Avalanching Photodetectors]{Measurements of the Quantum Yield of Silicon using Geiger-mode Avalanching Photodetectors}


\author*[1]{\fnm{Harry} \sur{Lewis}}\email{hlewis@triumf.ca}

\author[1]{\fnm{Mahsa} \sur{ Mahtab}}\email{m.mahtab.83@gmail.com}

\author[1]{\fnm{Fabrice} \sur{~Reti\`ere}}\email{fretiere@triumf.ca}

\author[1,2]{\fnm{Austin} \sur{De St. Croix}}\email{adestcroix@triumf.ca}

\author[1]{\fnm{Kurtis} \sur{Raymond}}\email{kraymond@triumf.ca}

\author[1]{\fnm{Maia} \sur{Henriksson-Ward}}\email{mhenrikssonward@triumf.ca}

\author[1,2]{\fnm{Nicholas} \sur{Morrison}}\email{19nem4@queensu.ca}

\author[1,3]{\fnm{Aileen} \sur{Zhang}}\email{azhang12@student.ubc.ca}

\author[1]{\fnm{Andrea} \sur{Capra}}\email{acapra@triumf.ca}

\author[1]{\fnm{Ryan} \sur{Underwood}}\email{runderwood@triumf.ca}

\affil*[1]{\orgname{TRIUMF}, \orgaddress{\street{4004 Wesbrook Mall}, \city{Vancouver}, \postcode{V6T 2A3}, \state{BC}, \country{Canada}}}

\affil[2]{\orgdiv{Department of Physics, Engineering Physics and Astronomy}, \orgname{Queen's University}, \orgaddress{\street{Stirling Hall, 64 Bader Lane}, \city{Kingston}, \postcode{K7L 3N6}, \state{ON}, \country{Canada}}}

\affil[3]{\orgdiv{Department of Engineering Physics}, \orgname{University of British Columbia}, \orgaddress{\street{Hennings Building, 6224 Agricultural Road}, \city{Vancouver}, \postcode{V6T 1Z1}, \state{BC}, \country{Canada}}}


\abstract{Accurate characterization of quantum yield is crucial to the reconstruction of energy depositions in silicon at the eV scale.  This work presents a new method for experimentally calculating quantum yield using vacuum UV-sensitive silicon photomultipliers (SiPMs), which can be used to determine the probabilities that a UV photon absorbed in a silicon crystal will produce one, two, or three electron-hole pairs. Results are presented which fully constrain the distribution at photon energies up to 7.75eV. This method works by exploiting the saturation of photon detection efficiency which occurs when these devices are biased sufficiently high above their avalanche breakdown voltage. The measured quantum yield values are lower than those that have been previously reported by experimental data and modelling - this is expected to impact the sensitivity of experiments searching for light dark matter through direct detection in semiconductors, and should also be taken into account when characterizing the performance of UV photodetectors with high quantum efficiency. Measurements have been taken using a Hamamatsu VUV4 and an FBK VUV-HD3 device, showing good agreements between devices, and at a range of temperatures from 163-233K. The validity of the method is assessed using supplementary measurements of absolute photon detection efficiency, and an additional novel method of measuring average quantum yield using DC current-voltage measurements of SiPMs is presented and used for corroboration.}

\maketitle
\section{Introduction}

It is known that, at sufficiently high photon energies, the absorption of a single photon in a semiconductor crystal may produce more than one energetic carrier pair \cite{hodgkinson_impact_1963},\cite{kolodinski_quantum_1993}. This occurs when the photon energy is such that, upon absorption, one or more of the charge carriers produced has energy greater than the threshold for impact ionization. The primary `hot' carrier will then distribute energy to other charges through impact ionization, resulting in the effective production of multiple carrier pairs \cite{hodgkinson_impact_1963}. The number of electron-hole pairs produced by such an event is referred to as the `quantum yield'. Understanding of this mechanism is of interest for the design and characterization of deep-UV photodetectors with very high quantum efficiency, such as novel black silicon devices \cite{zalewski_silicon_1983,canfield1998absolute,juntunen_near-unity_2016,tsang_quantum_2020,garin_black-silicon_2020}. It is also crucial for accurate event reconstruction of energy depositions in silicon at the eV scale, including the direct detection of light dark matter \cite{agnese_first_2018,sensei_collaboration_sensei_2019,essig_direct_2016,baehr_depfet_2017}.  Previous measurements of average quantum yield in silicon, performed using photodiodes with negligible internal losses, have shown that the average number of electron-hole pairs produced per photon increases with photon energy and starts to exceed unity at around $\sim$4 eV \cite{canfield1998absolute},\cite{korde_quantum_1987}. The probability distribution of the number of carrier pairs produced has also been modelled by Ramanathan \textit{et al.} \cite{ramanathan_ionization_2020}, but has not previously been experimentally verified.

In this paper, we introduce a new method for measuring quantum yield in silicon by leveraging the gain mechanisms which govern the operation of single photon avalanche diodes (SPADs) and SPAD arrays, which are referred to as silicon photomultipliers (SiPMs). This method permits the measurement of not only the average quantum yield at a given wavelength, but also the probabilities of producing different numbers of electron-hole pairs.

SPADs and SiPMs are avalanching photodetectors which are capable of single-photon sensitivity by operation in the `Geiger' mode, meaning that they are biased at a voltage greater than their breakdown voltage. In this case, the absorption of a single photon in the device and the associated production of an energetic carrier pair may trigger a diverging cascade of charge multiplication referred to as an avalanche breakdown event. This produces a measurable current pulse and permits the detection of single photons, and of photon counting in the case of SiPMs \cite{mcintyre_avalanche_1973}. The measurement of quantum yield using avalanching detectors has been made possible by the development of vacuum-UV sensitive SiPMs. These have been produced in the context of the nEXO experiment, which is a future liquid xenon TPC intended to search for neutrinoless double beta decay \cite{adhikari_nexo_2021}.

For a device operating in the Geiger mode \cite{renker_geiger-mode_2006}, the probability that a photon initiates an avalanche breakdown increases, according to a binomial distribution, with the number of electrons produced by the photon absorption. This is because a higher number of primary electrons entering the high-field region of the device permits several trials for triggering an avalanche. However, at a sufficiently high overvoltage (meaning the excess bias over the breakdown voltage that is applied to the device), the probability that a given electron entering the high-field region a SPAD will trigger an avalanche saturates to unity \cite{mcintyre_avalanche_1973}. This means that the overall probability of a photon absorption triggering an avalanche is insensitive to the quantum yield. The yield can then be measured by studying the rate at which the frequency of photon detection events approaches saturation with increasing overvoltage. In order to probe the quantum yield at a given wavelength the photon detection efficiency (PDE) as a function of overvoltage is compared to that at a reference wavelength, for which the yield is known to be one. A relatively higher PDE at low overvoltage indicates more trials to trigger an avalanche, and therefore a higher quantum yield. The probabilities that different numbers of electron-hole pairs are produced by a photon of given energy can be determined using binomial statistics, as detailed in Section \ref{section:QYtoPDE}.

For comparison, an analogous measurement has also been performed using average DC photocurrents. Current in the pre-breakdown linear mode is compared to that at saturated overvoltage, which determines the average quantum yield only.

The devices used in this work are candidates for incorporation into the nEXO photodetection subsystem, the Fondazione Bruno Kessler VUV-HD3 and the Hamamatsu VUV4 \cite{gallina_performance_2022}.
As the probability that holes trigger an avalanche does not reach saturation under the normal operating conditions of an SiPM, it is critical to ensure that all avalanches are electron-triggered. This is done by using p-on-n SiPMs, and wavelengths of light which have sufficiently short penetration depth to be fully absorbed in the p-doped region of the silicon. At 380 nm, the longest wavelength used in this study, 99.9\% of light is absorbed by a depth of 92.2 nm \cite{schinke_uncertainty_2015}. This is significantly shorter than the p-region depths calculated for these devices \cite{gallina_characterization_2019}, fulfilling the above condition.

\section{{Measurement apparatus}} 
\label{section:measurements}
The VERA (Vacuum, Efficiency, Reflection and Absorption) setup at TRIUMF is a dedicated apparatus for SiPM characterization. VERA is a substantially home-upgraded version of a Resonance TR-SES-200 transmissometer, which is now capable of optronics transmission, efficiency and reflection measurements under high vacuum (as low as 10\textsuperscript{-7} torr) and cryogenic conditions (as low as  -180\textdegree C). 

\begin{figure}[htp] 
\centering
\includegraphics[width=8cm]{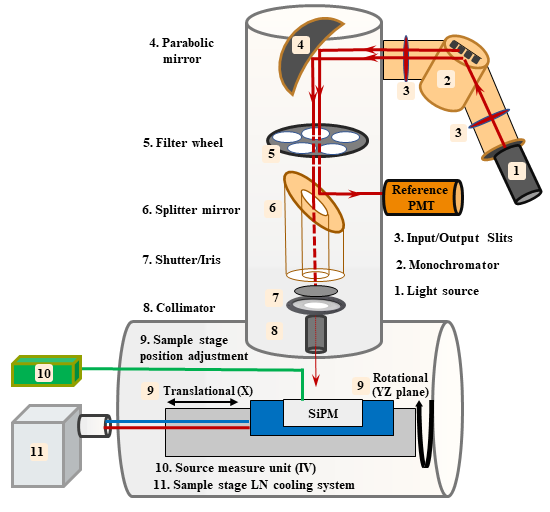}
\caption{Schematic showing the main components of the VERA setup. Figure shows the light source and monochromator, along with the beam conditioning path and the associated elements as numbered. The sample is mounted on an LN-cooled copper stage where it can be adjusted in the X-direction or rotated in the YZ-plane, to center the beam spot on the DUT and to change the incident beam angle, respectively.}
\label{VERA}
\end{figure}

As shown in Figure \ref{VERA}, the system is equipped with a deuterium lamp which provides a continuous spectrum from 140-830 nm, coupled to a VM200-monchromator. Data was taken with the monochromator's wavelength FWHM set to 9 nm. This is an experimental compromise as a smaller FWHM results in reduced optical power. The light beam is passed through a selectable filter wheel with an array of long-pass filters to suppress the grating-produced harmonics. The light beam is then split and reflected to a reference PMT (Hamamatsu R8486) that is used to track the light source flux and stability. The remaining part of the beam is passed through a shutter, iris and collimator assembly to adjust the beam spot size and shape. 

The sample stage is mounted on a motorized arm that can be adjusted to center and contain the light spot within the active area of the device, and to change the light angle of incidence by rotating the sample stage around the arm.

Each device is mounted to a custom-made PCB which is bolted onto the LN-cooled sample stage for optimal thermal contact and mechanical stability. The PCB includes a triax connector for DUT biasing and current readout, using a low noise triax cable which is connected to readout instrumentation through a ConFlat feedthrough. Current measurements were performed using a Keysight B2985A electrometer. For determining pulse rates at the waveform level, the output of each SiPM was converted to a voltage pulse using a low-noise, high speed transimpedance amplifier which was produced in-house. A CAEN DT5730 digitizer with a sampling rate of 500MHz and a dynamic range of 500mV was used to record the output pulses from the amplifier. In order to extract the rate of photon detection events, the effects of afterpulsing are accounted for by examining the time distribution of measured pulses. It is assumed that the rate of primary avalanches follows a Poisson distribution, and the mean time between events is extracted by applying an `unshadowing' transform to a histogram of the time differences between pulses. This method is described in detail in \cite{butcher_method_2017}. Time differences were extracted within a single waveform as the pulse rate was higher in all cases than the maximum waveform processing frequency of the digitizer. This meant that a minimum of 4 pulses per waveform were required, necessitating the use of waveforms up to 3ms in duration at lower pulse rates. When longer waveforms were required, fewer waveforms were taken and lower statistics were tolerated in order to complete the measurements in an acceptable timeframe, with a minimum of 35000 waveforms recorded and up to 100000 at higher pulse rates. A Keithley 6487 picoammeter/voltage source was used to bias devices for all measurements.

\section{Waveform-level measurements}
\subsection{Relating quantum yield to the PDE}
\label{section:QYtoPDE}
The photon detection efficiency (PDE, $\epsilon$) is the probability that a photon of wavelength $\lambda$ incident onto an SiPM generates an avalanche that is detected. The PDE depends on the following processes, which are assumed to be independent:
\begin{itemize}
    \item Fill factor (FF): The probability that the photon impinges onto a sensitive area. At normal incidence this corresponds to the proportion of the top surface of the device which comprises the active area of the diode pixels. 
    \item Transmittance ($T$): The wavelength-dependent probability that a photon passes through the passivation layer and enters the silicon. Some photons may be reflected at the various optical interfaces they encounter, including the silicon-passivation interface. Anti-reflective coatings are often used to minimize reflections through carefully tailored interference filters.
    \item Electron extraction from the surface ($P_x$): The wavelength-dependent probability that photo-electrons reach the region where the electric field is large enough to create an avalanche. Photo-generated electrons may be lost due to recombination when produced close to the surface if the distance between the surface and the depletion edge is significant relative to the minority carrier diffusion length. 
    We will assume a sharp transition for $P_x$ from zero probability of extraction below a certain depth, $D_x$, and 100\% extraction probability above $D_x$. This is supported by supplementary measurements of the absolute PDE of these devices, which are described in Section \ref{section:Psat}.
    \item The avalanche triggering probability ($P_a$): This work will use the model presented by Gallina \textit{et al.} \cite{gallina_characterization_2019}, for which the probability of avalanche is considered to be uniform for charge carriers produced in a given region of a SPAD. The wavelengths of light used in this study ensure that all photons are absorbed in the p-type region, and therefore only the electron triggering probability is considered. It is assumed that the probability of avalanche for a hole produced in the p-type region of a SPAD is negligible relative to that of an electron. Charge carriers which are produced close to the surface of the device and diffuse into the depletion region are also treated as if they had been produced in the p-type region.
    \item The quantum yield: We will label $\eta(\lambda)$ the average number of carrier pairs produced per photon. However, we introduce the probability ($p_n$) of producing $n$ electrons such that
\begin{equation}
    \eta(\lambda) = \sum_{n=1}^{n<\infty} p_n(\lambda)\cdot n 
     \label{eq:etasum}
\end{equation}

\end{itemize}

\noindent The PDE at overvoltage $V_o$ can then be written as follows, introducing $d(\lambda)$, the attenuation length in silicon of a photon with wavelength $\lambda$:
\begin{multline}
    \epsilon(V_o,\lambda) = T(\lambda) \cdot \mathrm{FF} \cdot \\\int_{0}^{\infty} \frac{e^{-z/d(\lambda) }}{d(\lambda)} \sum_{n=1}^{n<\infty} p_n(\lambda)\cdot[1-(1-P_x(\lambda) P_a(V_o))^{n})] dz
    \label{eq:pdeGen}
\end{multline}
where the sum term represents the overall probability of avalanche once the quantum yield and electron extraction probability is taken into account. This equation can be simplified using the approximation of a sharp transition from 0 to 1 of $P_x$ at $D_x$, and assuming that $d(\lambda)$ is significantly shorter than the distance from the device surface to the p-n junction.

\begin{multline}
        \epsilon(V_o,\lambda) = T(\lambda)\cdot \mathrm{FF} \cdot e^{-D_x/d(\lambda)}\cdot\\\sum_{n=1}^{n<\infty} p_n(\lambda)\cdot[1-(1-P_a(V_o))^{n}] 
    \label{eq:PDEgeneral}
\end{multline}
At sufficiently high bias voltage, the electron avalanche triggering probability $P_a$ reaches unity and the efficiency reaches a saturation value:

\begin{equation}
    \epsilon_{sat}(\lambda) = T(\lambda) \cdot \mathrm{FF} \cdot e^{-D_x/d(\lambda)} 
    \label{PDEsat}
\end{equation}
Using VERA, we measure the number of photons detected per second as a function of overvoltage, $R(V_o)$, for a given constant photon flux $\phi$. This can be expressed as:
\begin{equation}
    R(V_o) = \phi \cdot \epsilon_{sat}(\lambda) \cdot \sum_{n=1}^{n<\infty} p_n(\lambda)\cdot[1-(1-P_a(V_o))^{n}] 
    \label{eq:Rv}
\end{equation}
At sufficiently high bias voltage, $R(V_o) = R_{sat} = \phi \epsilon_{sat}$. We may then introduce the observable $\rho(V_0, \lambda)$ as follows:

\begin{equation}
    \rho(V_o,\lambda) = \frac{R(V_o)}{R_{sat}} = \sum_{n=1}^{n<\infty} p_n(\lambda)\cdot[1-(1-P_a(V_o))^{n}] 
    \label{eq:rhoVsum}
\end{equation}
This cancels all wavelength-dependent terms other than $p_n$. Thus, $p_n$ can be measured as a function of the wavelength, provided $P_a$ is known. For the SiPMs used in this study there is a range of wavelengths, between 350 and 400 nm, where $p_1$ is expected to be unity while still fulfilling the requirement that only electrons trigger avalanches. Therefore, $P_a(V_o)$ is measured at a reference wavelength in this range. $p_n$ is extracted for shorter wavelengths by fitting the data using equation \ref{eq:rhoVsum}. 

\subsection{Measurement and analysis technique}
\label{subsection:measanalysis}
\begin{figure}
    \centering
    \begin{subfigure}{\columnwidth}
     \caption{}
    \centering
    \includegraphics[width=3.5in]{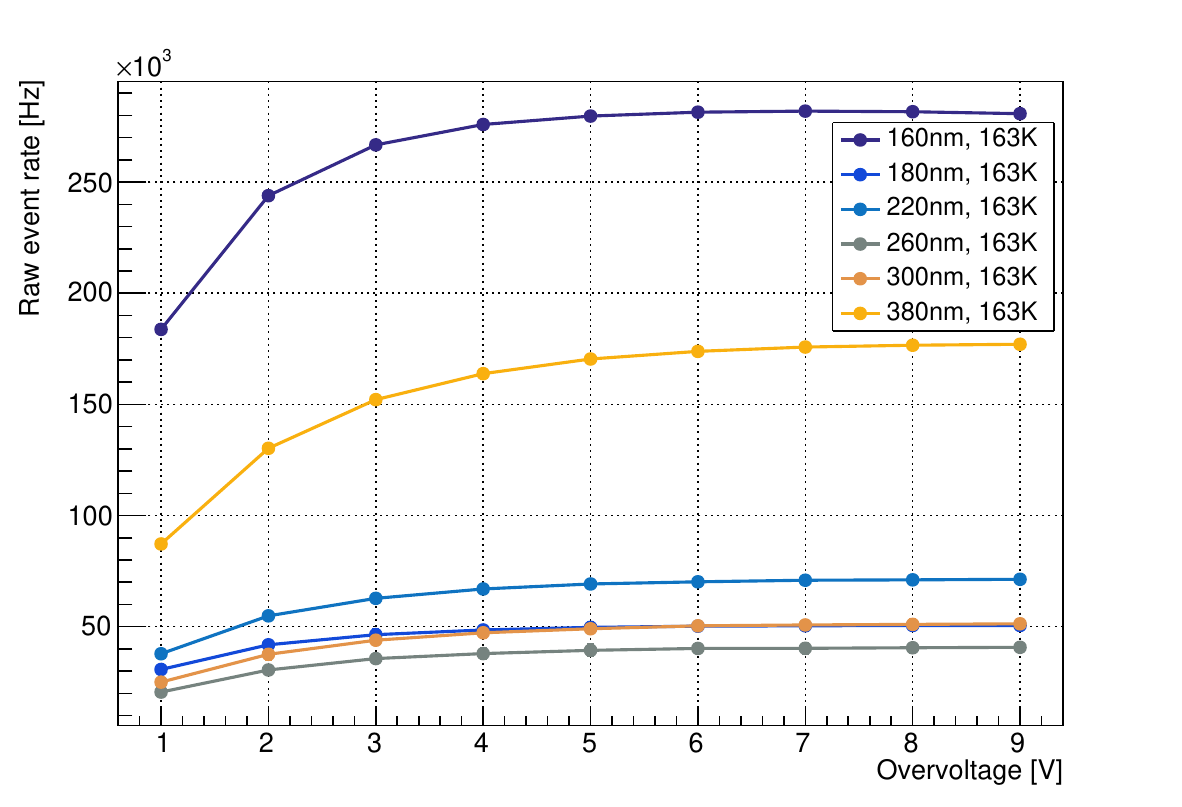}
    \end{subfigure} 
    \hfill
    \begin{subfigure}{\columnwidth}
    \centering
     \caption{}
    \includegraphics[width=3.5in]{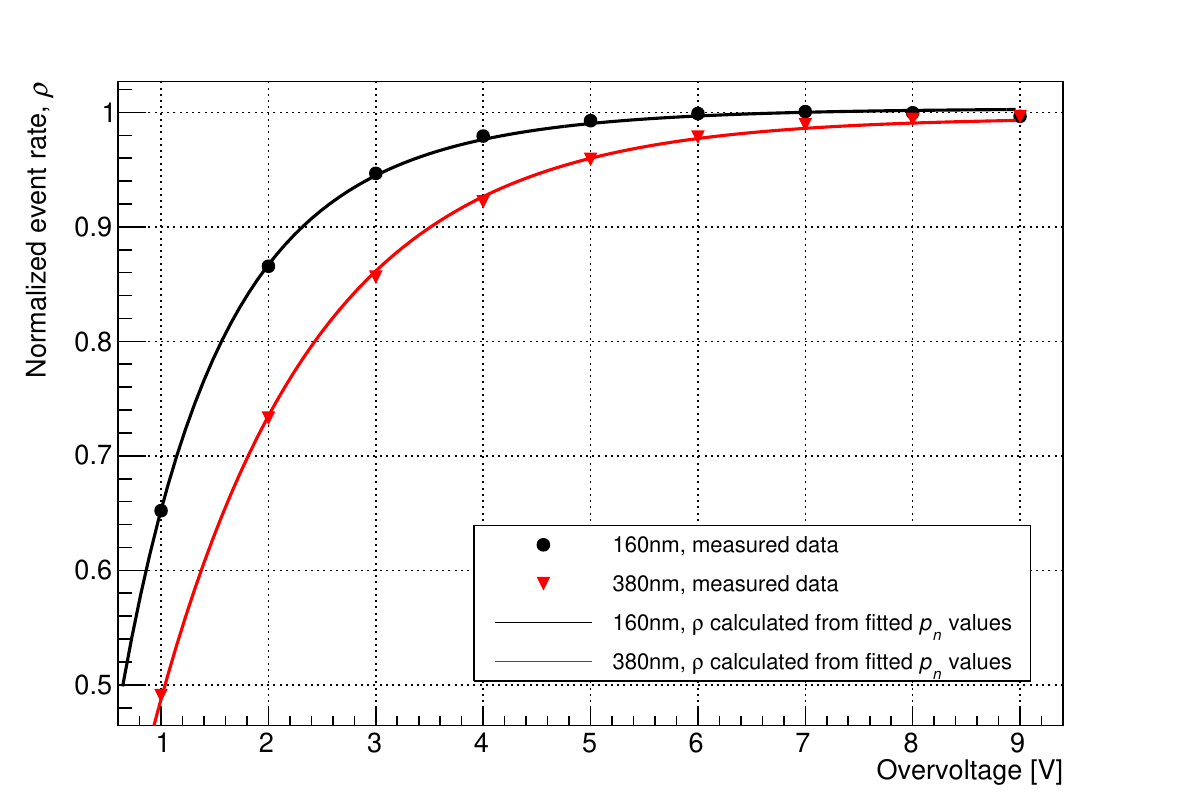}
    \end{subfigure}

    \caption{(a) Raw photon detection rates for the Hamamatsu VUV4 at 163 K. The most significant factor in the variation in pulse rate is the different optical power available at each wavelength. (b) Rates for 160 and 380 nm normalized to their saturation values, shown with modelled data calculated using the fitted $p_n$ values.}
    \label{fig:eventrates}
\end{figure}

The rate of current pulses under various illumination and bias conditions was measured for each SiPM device by recording waveforms under each operating condition as described in section \ref{section:measurements}. 
The measured photon detection rates for various wavelengths at 163 K are shown in figure \ref{fig:eventrates}(a). Saturation is indicated by the flattening behaviour of the curves at high overvoltage. 

$R_{sat}$ is determined by fitting the event rates at a given wavelength to an empirical function of the form $A\cdot(1-e^{-Va})$, and determining the asymptotic maximum of this function. Normalizing event rates to $R_{sat}$ yields $\rho(V)$ as shown in figure \ref{fig:eventrates}(b) for 163 K. A rate which increases more quickly to saturation, as observed at 160 nm, indicates a higher quantum yield. 

\begin{figure}
    \centering
    \includegraphics[width=3.5in]{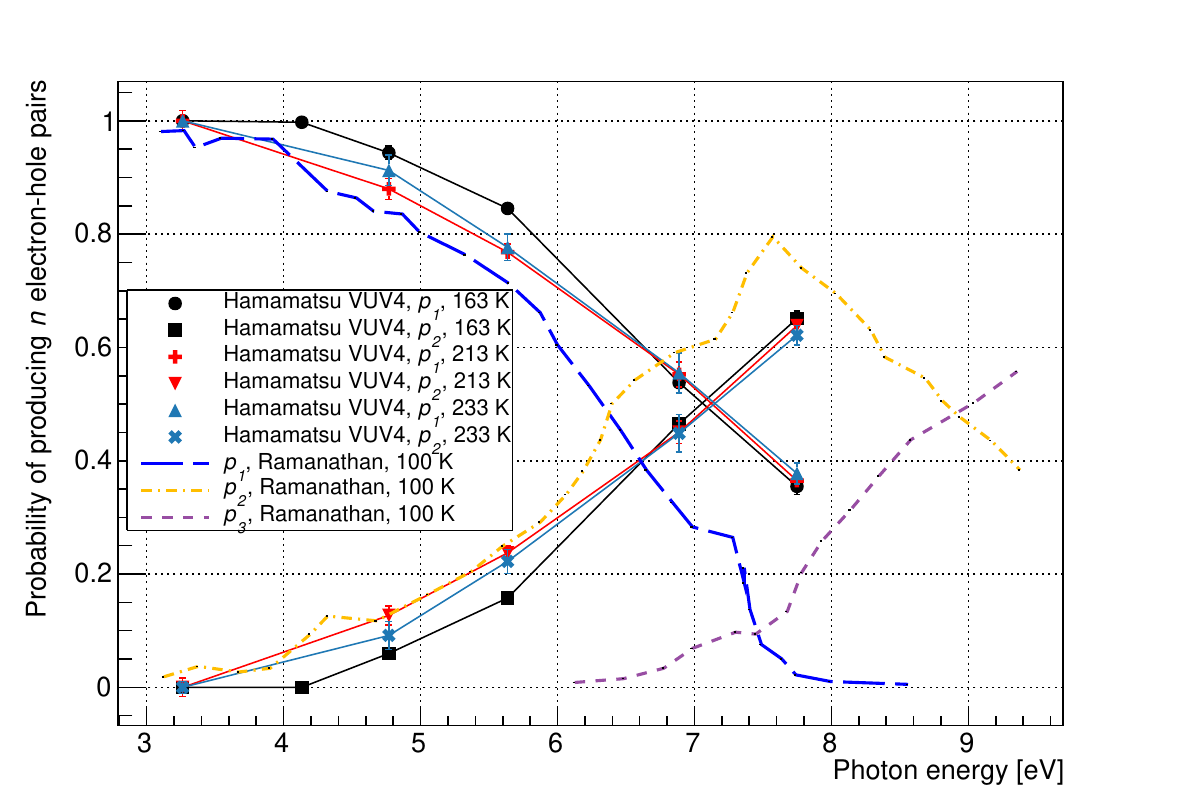}

        \caption{Calculated values for $p_1$ and $p_2$ as a function of photon energy, shown for the Hamamatsu VUV4 device at various temperatures. Modelled data from Ramanathan \textit{et al.} \cite{ramanathan_ionization_2020} are shown for comparison.}
    \label{fig:probabilities}
\end{figure}

\begin{figure}
    \centering
    \centering
    \includegraphics[width=3.5in]{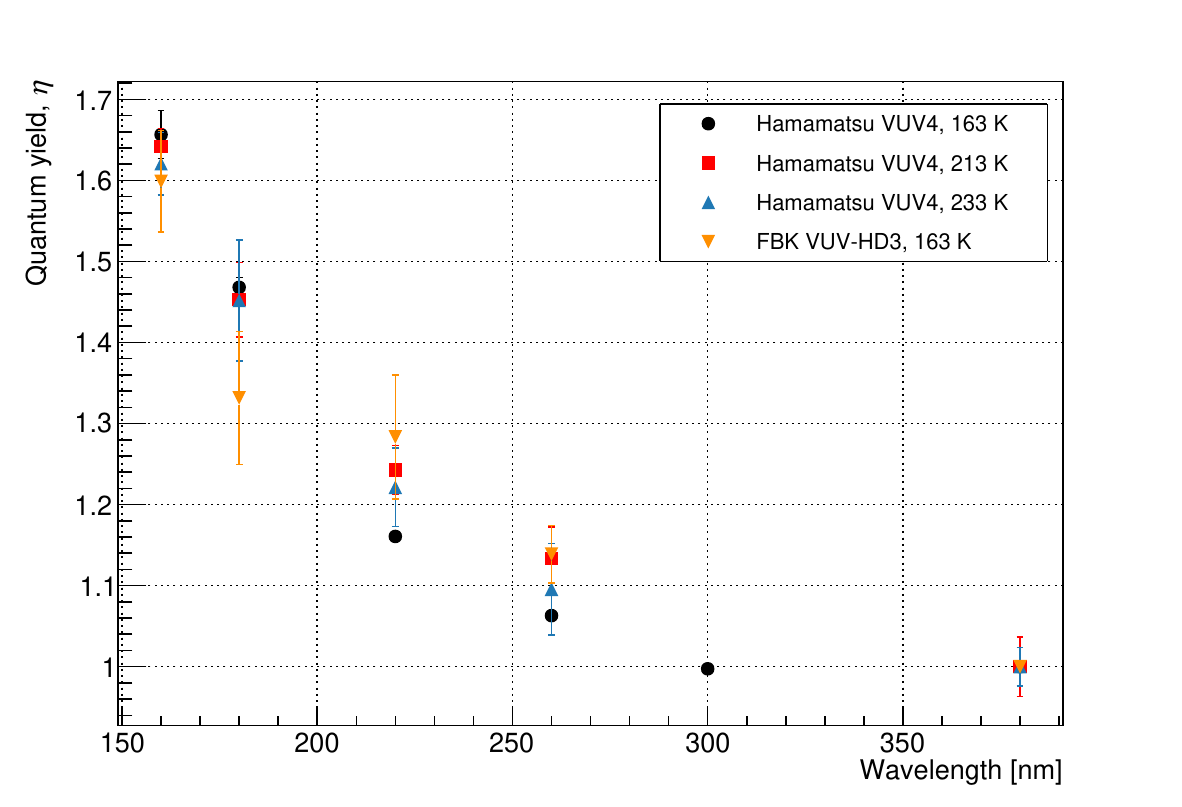}
    \caption{Values of $\eta$ as a function of wavelength, calculated using waveform-level measurements for the Hamamatsu VUV4 and FBK VUV-HD3. Data for the VUV4 are given for a range of temperatures.}
    \label{fig:waveformEta}
\end{figure} 

It is then possible to measure the probability distribution of the quantum yield by referring to equation \ref{eq:rhoVsum} and performing a fit for $p_n$ at different values of $n$. $P_a(V)$ is set to the value of $\rho(V)$ at 380 nm. 

The fit was first performed for three parameters: $p_1$, $p_2$ and $p_3$. It is assumed that photons in the wavelength range under consideration do not have sufficient energy to produce four carrier pairs. With all parameters limited between 0 and 1, the resulting fit gave values $<1.5\times10^{-6}$ for $p_3$ at all wavelengths.  These values are several orders of magnitude smaller than those for $p_1$ and $p_2$ and are considered to correspond to a fit result of zero for $p_3$. The fit was therefore repeated using only two parameters, $p_1$, and $p_2$.  The resulting values for $p_1$ and $p_2$ were identical to those produced when $p_3$ was also incorporated. Reduced $\chi^2$ values for the fits varied from 0.32 (VUV4, 233K, 220nm) to 1.92 (VUV-HD3, 163K, 220nm).
The results of this fit indicated that the probability of a photon in this wavelength range producing three carrier pairs is negligible. The fitted values for $p_1$ and $p_2$ are shown in figure \ref{fig:probabilities} as a function of photon energy. The modelled $p_n$ values given by Ramanathan \textit{et al.} \cite{ramanathan_ionization_2020} are shown for comparison, with our measured data indicating lower quantum yield for a given energy. Measurements were taken at 163~K for the FBK device, and at 163, 213, and 233~K for the Hamamatsu device. Measurements showed agreement between the devices within our estimated error and no temperature dependence was observed. $\eta$ was determined by taking a weighted average of the $p_1$ and $p_2$, as shown in equation \ref{eq:etasum}, with the calculated values shown in figure \ref{fig:waveformEta}. The relative nature of this measurement minimizes systematic error provided that the assumptions of the model are correct. This is because the only operating condition varied between data points at a given wavelength is the bias voltage, with all other conditions affecting the rate of SPAD pulses considered to remain invariant. The operating conditions most likely to affect pulse rate are the optical power on the device and the temperature, which are respectively controlled by monitoring the photon flux in the setup using a VUV-sensitive PMT and controlling the temperature to within 0.1\textdegree C using a feedback system. As such we consider that the systematic errors in this measurement are negligible compared to the statistical error. These assumptions are addressed further in section \ref{section:Psat}. The statistical error of the measurement increases with temperature due to higher dark counts in the SiPMs.

\section{DC current measurements}

\begin{figure}
    \centering
    \includegraphics[width=3.5in]{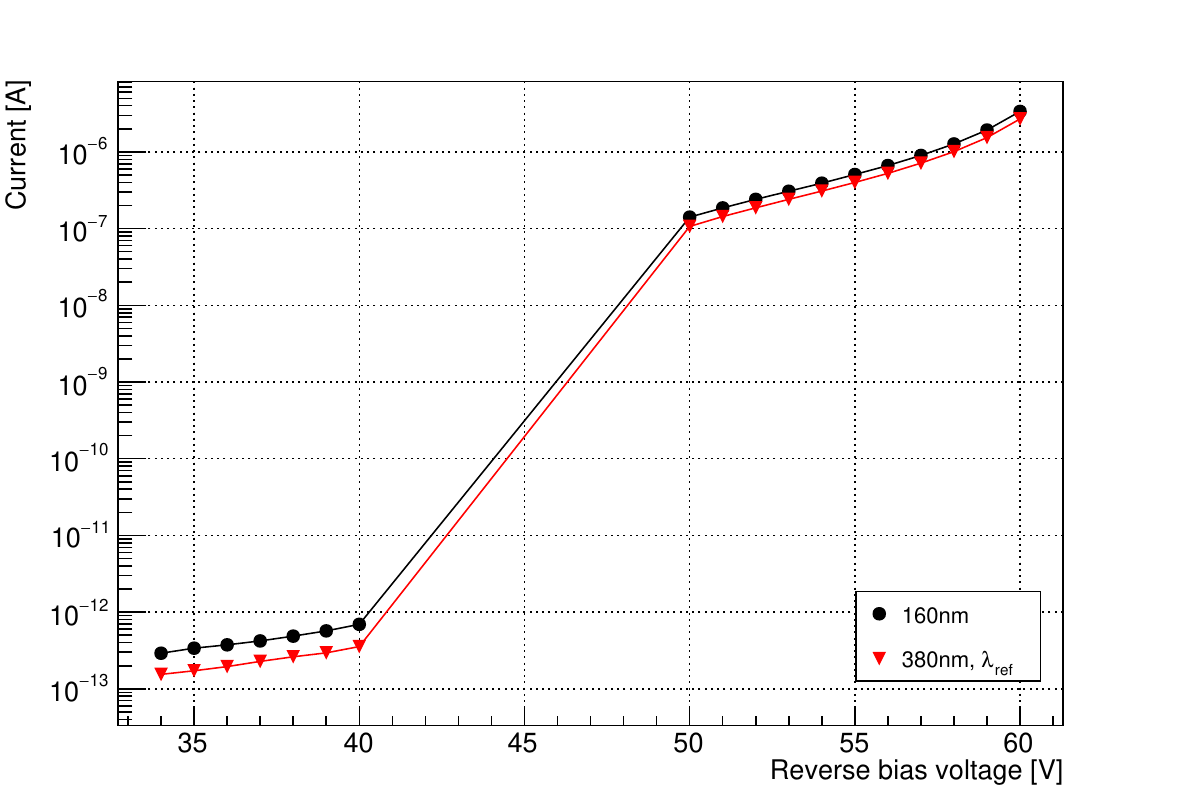}
        \caption{A typical IV characteristic for the Hamamatsu VUV4, here showing data for 160 nm and 380 nm. Data were taken in the linear mode (left) and saturation mode (right), with bias voltages around breakdown omitted.}
    \label{fig:IVcurve}
\end{figure}

It is also possible to make a similar measurement of $\eta$ using the average DC output of an SiPM, recorded with a picoammeter. The data taking and analysis are straightforward in this case, but the measurement of very low currents is required. This method can be used to calculate the average quantum yield only, and primarily serves to corroborate the results of the previous section. Measurements were performed using the Hamamatsu VUV4 device only, as similar behaviour between devices is expected given the results shown in figure \ref{fig:waveformEta}. The DC current output of the SiPM is measured under two different bias conditions: 1) at sufficiently high bias voltage that the device PDE is saturated, and 2) at low bias voltage where the SiPM is operating with a non-Geiger mode linear gain. Example data for each mode are shown in figure \ref{fig:IVcurve}. The linear-mode gain value is wavelength-independent for a given bias, provided that the wavelength is short enough that the vast majority of impact ionization events are electron-initiated. This is the case if all incident photons are absorbed in the p-type layer of the device, which is true for all wavelengths used in this study. 

In the linear mode, following the assumptions of the previous section, the average current $I_p$ at bias voltage $V$ can be written as follows:
\begin{equation}
    I_{l,tot}(V) = I_{l,d}(V) + \phi \cdot T(\lambda) \cdot \mathrm{FF} \cdot e^{-D_x/d(\lambda)} \cdot M(V) \cdot\eta(\lambda)
\end{equation}
where $I_{l,d}(V)$ is the dark current and $M(V)$ is the linear gain. This current is proportional to $\eta$. $I_l$ is the linear mode photocurrent  $I_{l,tot} - I_{l,d}$. Taking the ratio of linear mode photocurrent (using a constant linear-mode bias voltage) at a given wavelength to that at a reference wavelength, $\lambda_{ref}$, where $\eta(\lambda_{ref}) = 1$, then gives:

\begin{equation}
\frac{I_l(\lambda)}{I_l(\lambda_{ref})} = \frac{\phi(\lambda) \cdot T(\lambda) \cdot e^{-D_x/d(\lambda)} \cdot \eta(\lambda)}{\phi(\lambda_{ref}) \cdot T(\lambda_{ref}) \cdot e^{-D_x/d(\lambda_{ref})} \cdot \eta(\lambda_{ref})}
\label{eq:ilinratio}
\end{equation}
It is then possible to determine $\eta(\lambda)$ by considering the current in the saturation mode:

\begin{multline}
    I_{s,tot}(V) = I_{s,d}(V) + \phi \cdot T \cdot \mathrm{FF} \cdot e^{-D_x/d(\lambda)} \cdot Q_{a}(V) 
\end{multline}
where $I_{s,d}(V)$ is the dark current and $Q_a(V)$ is the charge produced per avalanche, including additional avalanches produced by cross-talk and after-pulsing. Both quantities depend on bias voltage. $I_s$ is the saturation mode photocurrent  $I_{s,tot} - I_{s,d}$. This current does not depend on $\eta$, and $Q_{a}$ is independent of wavelength. It is therefore possible to cancel all wavelength-dependent terms other than $\eta$ by considering the ratio $\frac{I_s(\lambda)}{I_s(\lambda_{ref})}$ of currents in the saturation mode at a constant bias voltage:

\begin{equation}
\frac{I_s(\lambda)}{I_s(\lambda_{ref})} = \frac{\phi(\lambda) \cdot T(\lambda) \cdot \mathrm{FF} \cdot e^{-D_x/d(\lambda)}}{\phi(\lambda_{ref}) \cdot T(\lambda_{ref}) \cdot \mathrm{FF} \cdot e^{-D_x/d(\lambda_{ref})}}
\label{eq:isatratio}
\end{equation} 
Equations \ref{eq:ilinratio} and \ref{eq:isatratio} can then be combined to give:

\begin{equation}    \eta(\lambda) = \sfrac{\frac{I_l(\lambda)}{I_l(\lambda_{ref})}}{\frac{I_s(\lambda)}{I_s(\lambda_{ref})}}
\label{eq:doublecurrentratio}
\end{equation}

\begin{figure}
    \centering
    \includegraphics[width=3.5in]{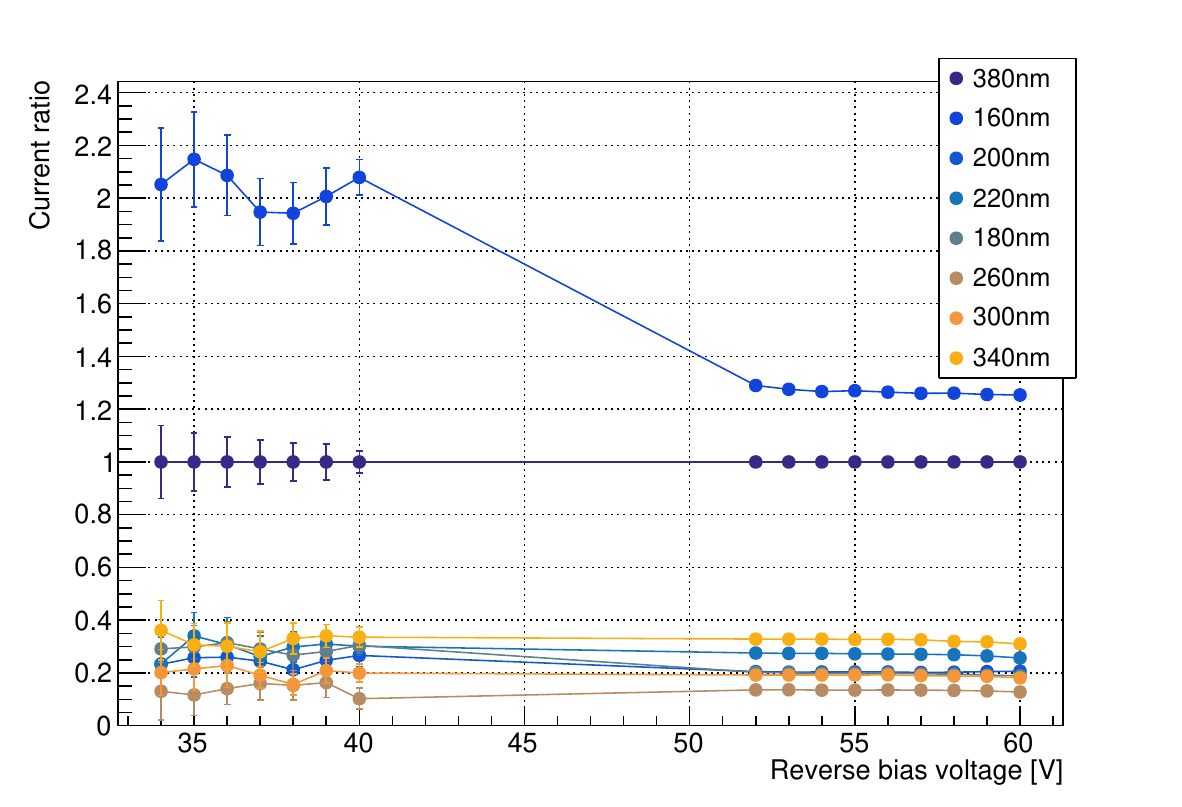}

    \caption{Current ratios $\frac{I(\lambda)}{I(\lambda_{ref})}$ at various bias voltages. Bias at 34-40 V indicates linear mode operation, and bias at 52-60 V indicates saturation mode operation.}
    \label{fig:currentratios}
\end{figure}

\begin{figure}
    \centering
    \includegraphics[width=3.5in]{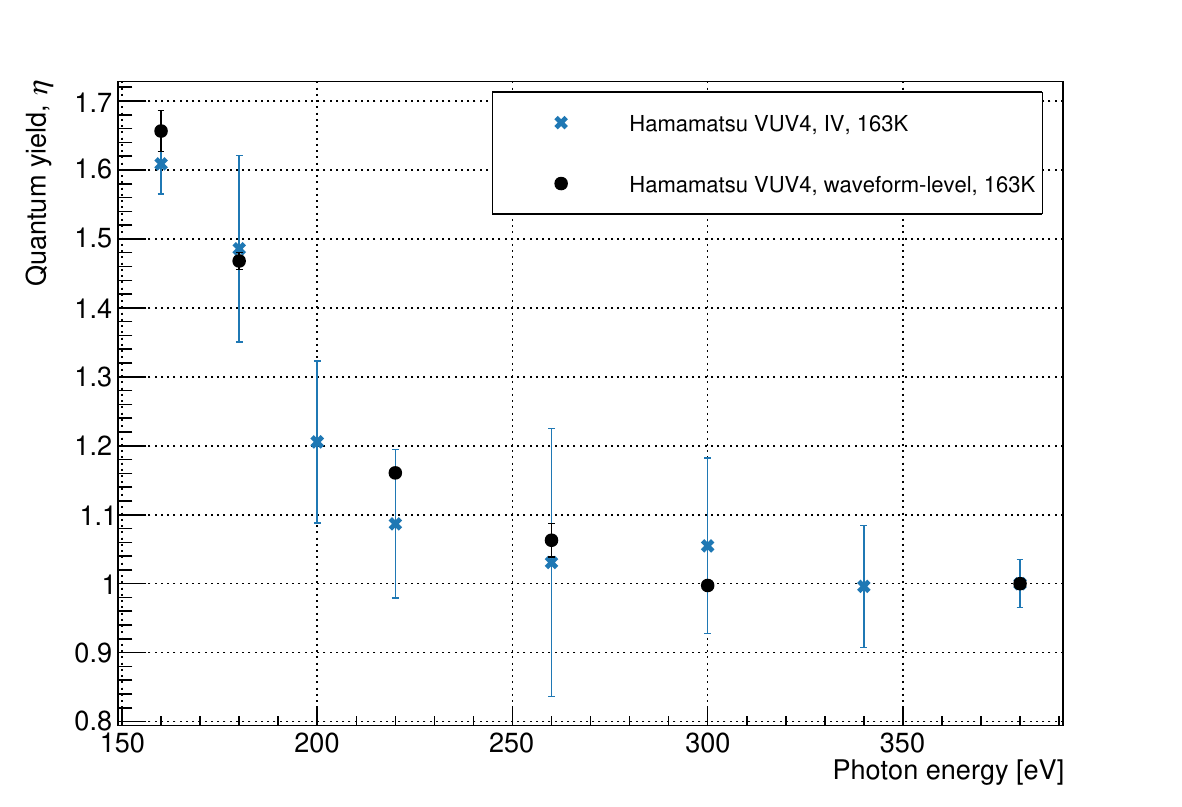}
    \caption{Values for $\eta$ as a function of wavelength, calculated using waveform-level and IV measurements.}
    \label{fig:etaIV}
\end{figure}

The ratios $\frac{I_l(\lambda)}{I_l(\lambda_{ref})}$ and $\frac{I_s(\lambda)}{I_s(\lambda_{ref})}$ are independent of bias voltage, and any given linear mode bias voltage can be combined with any given saturation mode bias voltage without affecting the validity of equation~\ref{eq:doublecurrentratio}. This allows an average value for each of these two ratios to be calculated from a range of bias voltages in each mode, increasing statistical precision. The different ratios measured at a range of bias voltages and wavelengths are shown in figure \ref{fig:currentratios}. The calculated $\eta$ values are included in figure \ref{fig:etaIV} and agree with those determined using the waveform-level measurements described above.

\section{Assessment of model validity}
\label{section:Psat}
\begin{figure}
    \centering
    \includegraphics[width=3.5in]{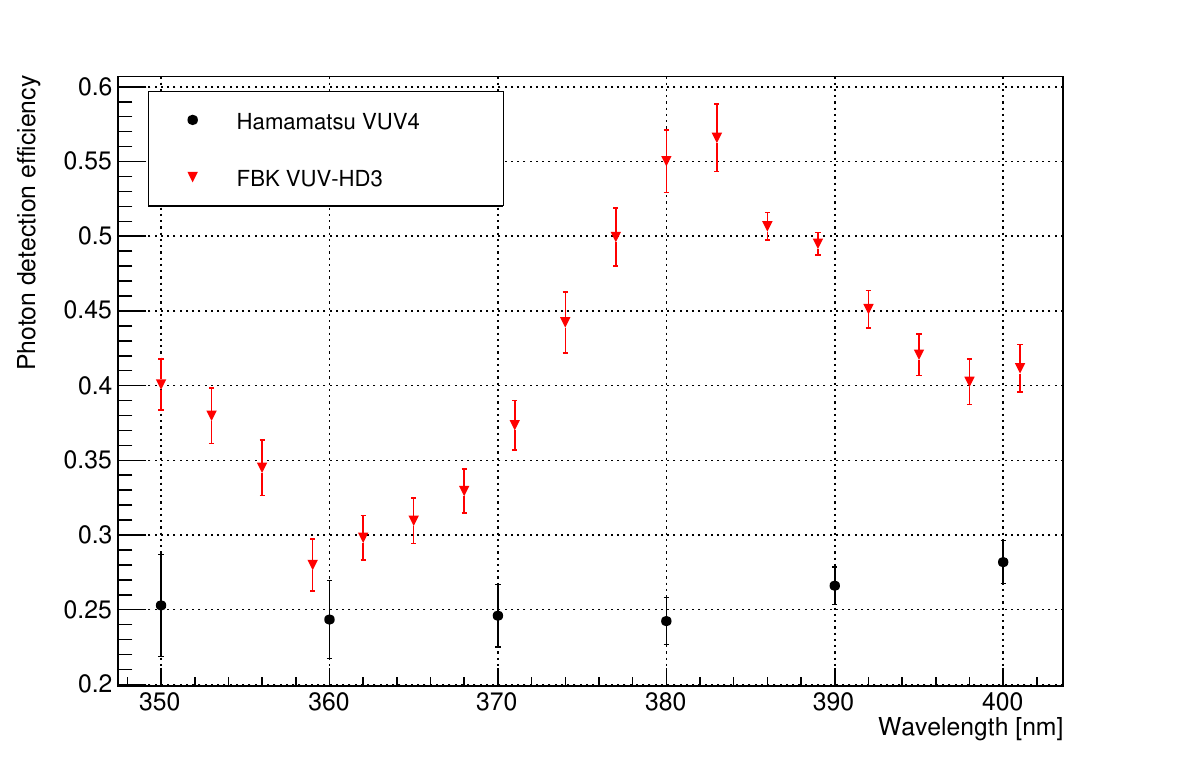}
    \caption{PDE values at 7V overvoltage for the devices used in this study, in the wavelength range 350-400 nm.}
    \label{fig:PDE}
\end{figure}

The PDE values of the devices used in this study have also been measured at a range of wavelengths using the VERA setup. The measurement was performed by comparing the DC current in the device at various wavelengths to current measured by a NIST-calibrated photodiode under the same optical flux. The DC responsivity of the device was converted to PDE by measuring the gross charge per photon detection event using waveform-level measurements at each overvoltage, with the calculated values shown in figure \ref{fig:PDE}. Data were measured for a wavelength range of 350-400nm at 7V overvoltage, which is here assumed to be sufficient to saturate the PDE.

\begin{figure}
    \centering
    \begin{subfigure}{\columnwidth}
     \caption{}
    \centering
    \includegraphics[width=3.5in]{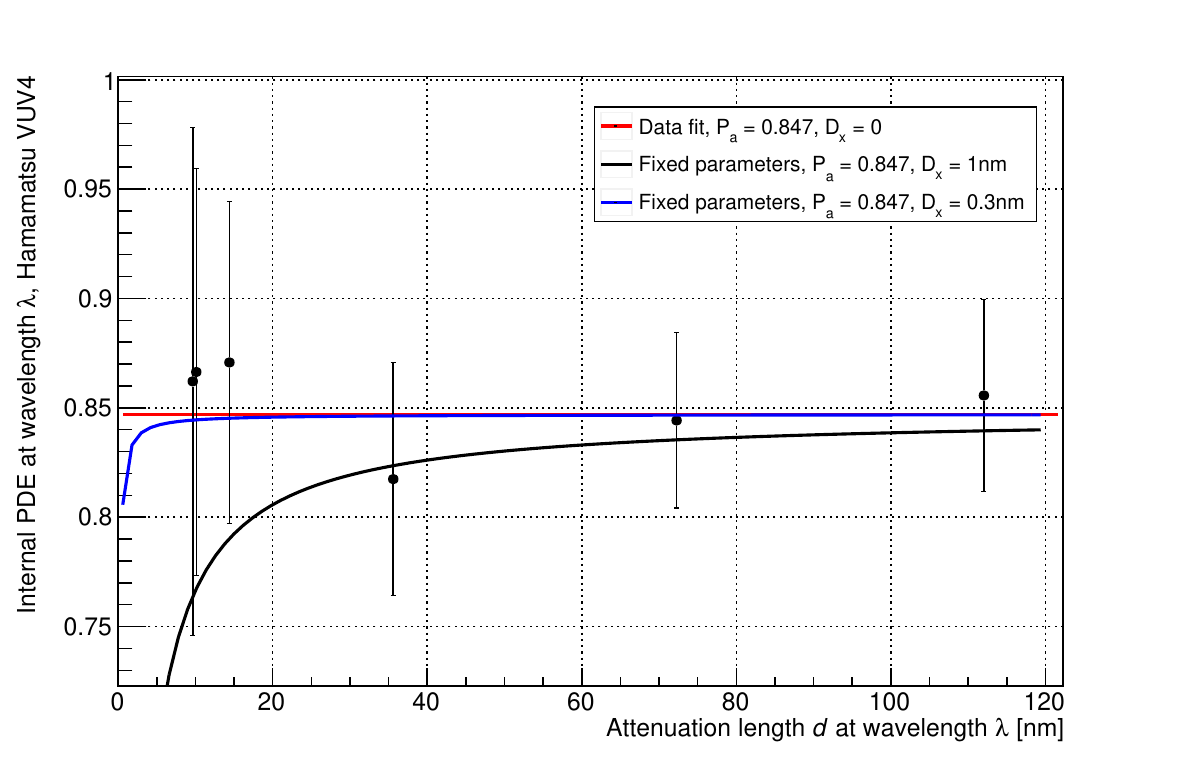}
    \end{subfigure} 
    \hfill
    \begin{subfigure}{\columnwidth}
    \centering
     \caption{}
    \includegraphics[width=3.5in]{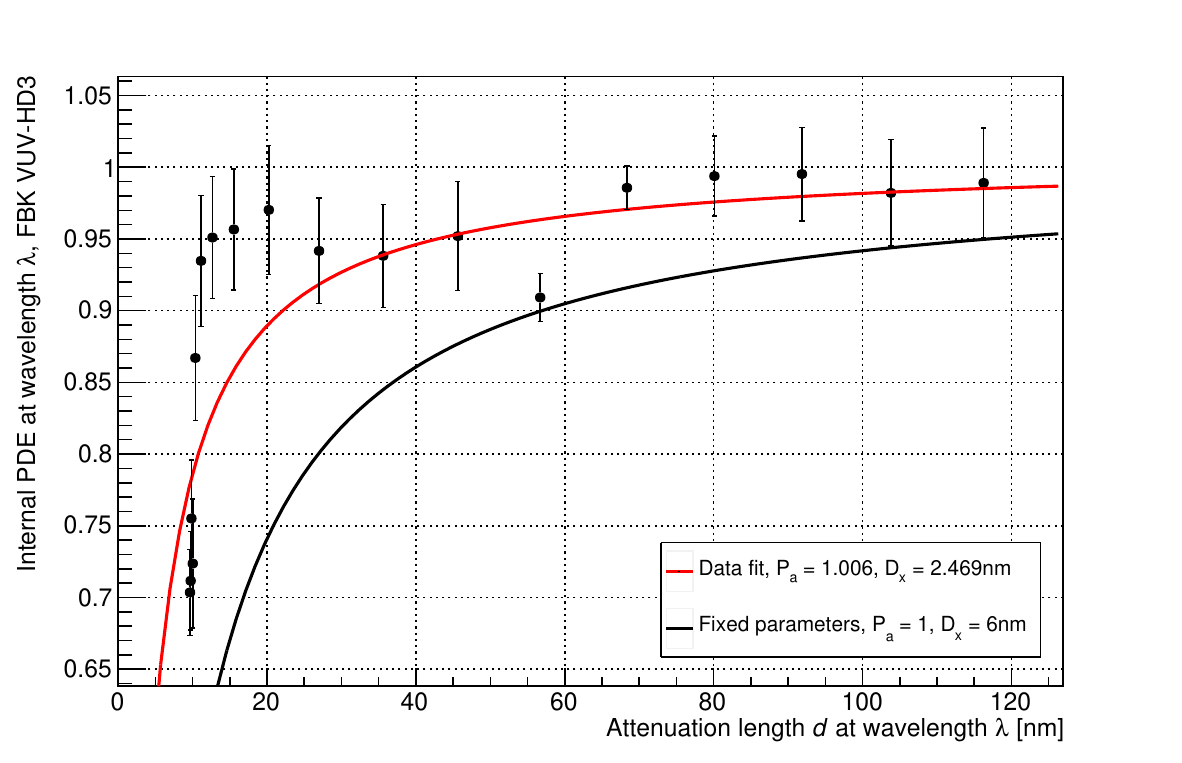}
    \end{subfigure}

    \caption{(a) Internal PDE, $\epsilon/(T \cdot \mathrm{FF})$, at wavelength $\lambda$ against $d(\lambda)$ for the Hamamatsu VUV4, shown with fitted and fixed-parameter functions of the form $P_{sat} \cdot e^{-D_x/d}$. Various values of $D_x$ are shown to indicate that this parameter is not fully constrained by the measured data. (b) Internal PDE at wavelength $\lambda$ against $d(\lambda)$ for the FBK VUV-HD3, shown with a fitted function of the form $P_{sat} \cdot e^{-D_x/d}$.}
    \label{fig:PDEovertransmission}
\end{figure}

Calculations of $P_a$ using the measured PDE values and modelling of the transmission of photons into the device suggest that $P_a$ may saturate at a value less than unity in the Hamamatsu VUV4. This is in contradiction with existing theory \cite{mcintyre_avalanche_1973}\cite{mcintyre_new_1999}, and this discrepancy may be due to errors in other parameters such as the manufacturer-provided fill factor of the device. We will, however, consider the effects on our calculated quantum yield values of a saturated $P_a$ less than unity, denoted as $P_{sat}$. This section will also evaluate the assumption that the probability of electron extraction from the device surface transitions sharply from 0 to 1 when photons are absorbed at a depth of $D_x$ from the surface. It should be noted that $D_x$ is an effective parameter intended to permit an empirical quantification of the probability of extraction, and does not correspond to the physical device structure. In the case that $P_a$ saturates to $P_{sat}$, equation \ref{eq:Rv} becomes:

\begin{multline}
    R(V)  = \phi \cdot T(\lambda) \cdot \mathrm{FF} \cdot e^{-D_x/d(\lambda)} \cdot \\\sum_{n=1}^{n<\infty} p_n(\lambda)\cdot [1-(P_{sat}\cdot\rho_{ref}(V_o))^{n}] 
    \label{eq:lowPa}
\end{multline}
Where $\rho_{ref}(V_o)$ is $\frac{R(V_o)}{R_{sat}}$ at a reference wavelength where $\eta = 1$. The value of $P_{sat}$ was determined using the measured PDE in the wavelength range 350-400 nm, where $\eta = 1$ and all avalanches are electron-triggered. In this case equation \ref{eq:PDEgeneral} becomes:

\begin{equation}
        \epsilon_{sat}(\lambda) = T(\lambda) \cdot \mathrm{FF} \cdot e^{-D_x/d} \cdot P_{sat}
\end{equation}
$P_{sat}$ and $D_x$ can then be determined knowing $T$, $FF$, and $d$. $T$ is wavelength dependent and was calculated using Fresnel equations based on the thicknesses of the SiO$_2$ passivation layers on each SiPM device. Optical constants for SiO$_2$ were taken from \cite{lithography}. Values of these thicknesses were taken from \cite{sipmRef_nexo_guofu} and \cite{raymond_stimulated_TED_2024} for the Hamamatsu and FBK devices respectively. Values of $d$ were calculated using optical constants for silicon as given in \cite{nSi_aspnes}. Values of $FF$ are given by the device manufacturers as 0.6 for the VUV4 and 0.8 for the VUV-HD3. Data for the internal PDE of each device, $\epsilon/(T \cdot \mathrm{FF})$, were fitted to a function of the form $P_{sat} \cdot e^{-D_x/d}$ as shown in figure \ref{fig:PDEovertransmission}(a) for the VUV4 and \ref{fig:PDEovertransmission}(b) for the VUV-HD3. This yielded values for $D_x$ and $P_{sat}$ as shown in table \ref{table:PaDx}.
 Data for the FBK device yielded a $P_{sat}$ value very close to unity, which is considered to justify the assumption that $P_a$ saturates to unity in this device. $D_x$ for this device was evaluated as 2.469 nm. However the fit to the data was poor, with a reduced $\chi^2$ value of 3.33. This indicates that our model of internal PDE is not consistent with this data, which is likely due to error in the optical transmission calculations used
 
 to determine the internal PDE from the absolute PDE. This is supported by the oscillation patterns visible in the data in figure \ref{fig:PDEovertransmission}(b), which appear in the absolute PDE data in figure \ref{fig:PDE} but should be cancelled by an accurate transmission calculation. As such, modelled data using fixed parameters are compared to the measured internal PDE data in order to provide an upper bound for $D_x$. Figure \ref{fig:PDEovertransmission}(b) shows modelled data using a fixed $D_x$ of 6nm, and we consider that larger values of $D_x$ cannot be consistent with the measured data in this wavelength range. As such we take 6nm to be the upper bound for $D_x$ in this device.
The Hamamatsu device shows no significant decrease in internal PDE for wavelengths with shallower penetration into the device. The lack of available PDE data at very short wavelengths means that $D_x$ cannot be well constrained for this device using our model for internal PDE. It is, however, assumed that the internal PDE will drop off at wavelengths with arbitrarily short \textit{d}. This is because the electric field in a SPAD pixel cannot extend to the device surface while maintaining the device functionality as a diode. As described for the FBK device, we have constrained the upper bound of $D_x$ by comparing the measured internal PDE data to modelled data using fixed parameters. Figure \ref{fig:PDEovertransmission}(a) shows modelled data for $D_x$ values of 0.3nm and 1nm. 1nm is the maximum value of $D_x$ which can be used to produce modelled internal PDE data compatible with the measured data, and is considered to be the upper bound for $D_x$ in this device. The upper bounds for $D_x$ for both devices are significantly lower than the attenuation lengths in silicon for the wavelengths of light used in this study. This indicates that internal losses due to recombination are very low in these devices, and supports the validity of the assumption given in equation \ref{eq:PDEgeneral}. A model incorporating a linear increase in $P_x$ over a region close to the surface of the device, rather than a sharp transition from 0 to 1, was also investigated and showed negligible effect on the final calculated values of quantum yield.

\begin{figure}
    \centering
    \includegraphics[width=3.5in]{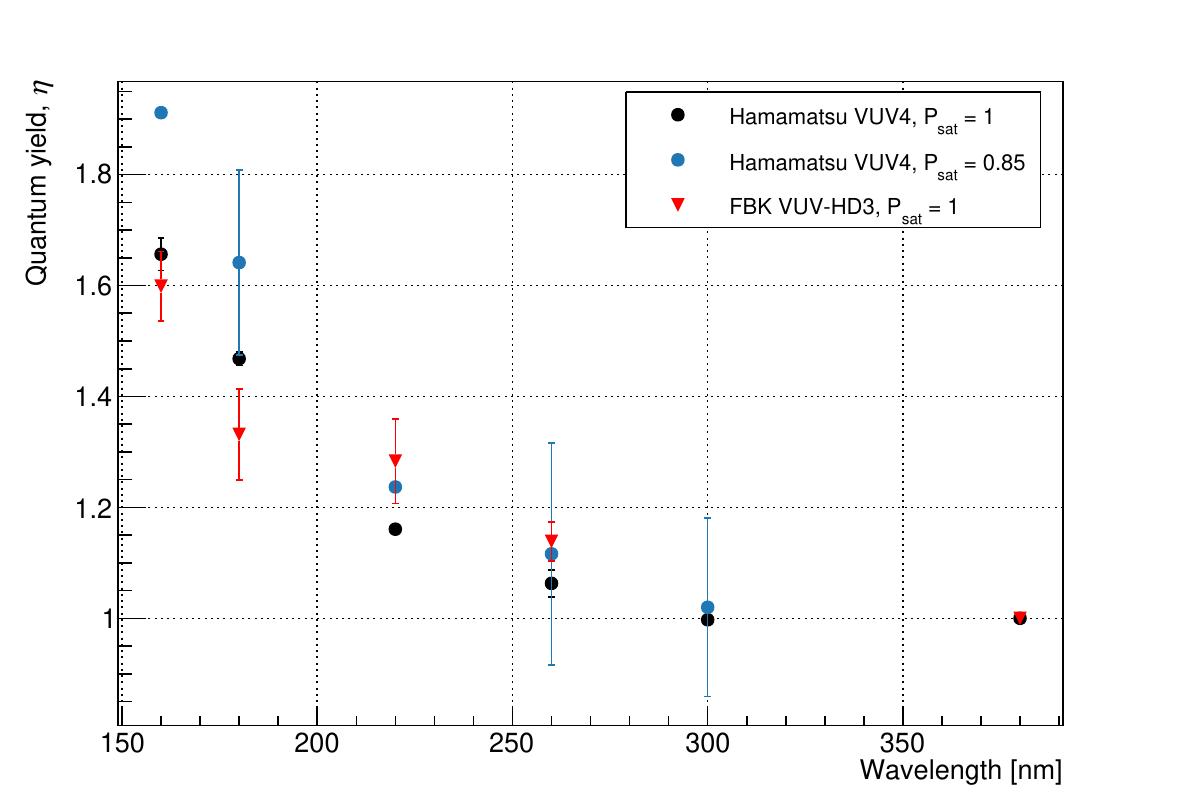}
    \caption{Values for $\eta$ calculated using waveform-level measurements, showing the effect of a $P_{sat}$ value of 0.85 for the Hamamatsu VUV4.}
    \label{fig:QY with Psat}
\end{figure}

The value of $P_{sat}$ for the Hamamatsu device is, however, well constrained at approximately 0.85. As stated above, it is likely that this discrepancy is due to errors in the calculated transmission or provided fill factor, or to border effects which may reduce the PDE of the device at the edges of the active area. The effects of $P_a$ saturating to this value were examined by fitting the raw measured event rates for each wavelength to equation \ref{eq:lowPa} for $p_n$ at $n = 1,2,3$. The overall value of $ \phi \cdot T \cdot \mathrm{FF} \cdot e^{-D_x/d}$ was used as an additional fitting parameter, resulting in increased error compared to the method used in Section \ref{subsection:measanalysis}. The resulting values of $\eta$, calculated using equation \ref{eq:etasum}, are shown in figure \ref{fig:QY with Psat}. The reduced $P_{sat}$ results in slightly higher calculated values for $\eta$.

\begin{table}[h!]
    \centering
    \caption{Values of $P_{sat}$ and $D_x$ for devices used in this study}
    \begin{tabular}{l c r}
        \toprule
        Device & $P_{sat}$ & $D_x$ [nm] \\
        \midrule
        Hamamatsu VUV4 & 0.847$ \pm 0.023$ & $<1$ \\
        FBK VUV-HD3 & 1.006$ \pm 0.011$ & $<

        6$ \\
        \bottomrule
    \label{table:PaDx}
    \end{tabular}
\end{table}

\section{Discussion}

We have demonstrated a new method of measuring the probability distribution of the quantum yield of silicon, which exploits the saturation of PDE in Geiger-mode avalanching photodetectors. This method can determine the probabilities that a photon will produce one, two, or three electron-hole pairs upon absorption, with the distribution fully constrained up to a photon energy of 7.75eV. The SiPM devices used in this study are sensitive to wavelengths as low as 125 nm, meaning that quantum yield could be measured up to energies of nearly 10 eV with an appropriate light source. Developing SPADs sensitive to shorter wavelengths is challenging due to the lack of available passivation materials capable of transmission below 120nm, meaning that the quantum yield of photon energies in the `UV-gap' of 10-50eV remains difficult to probe.
\begin{figure}
    \centering
    \includegraphics[width=3.5in]{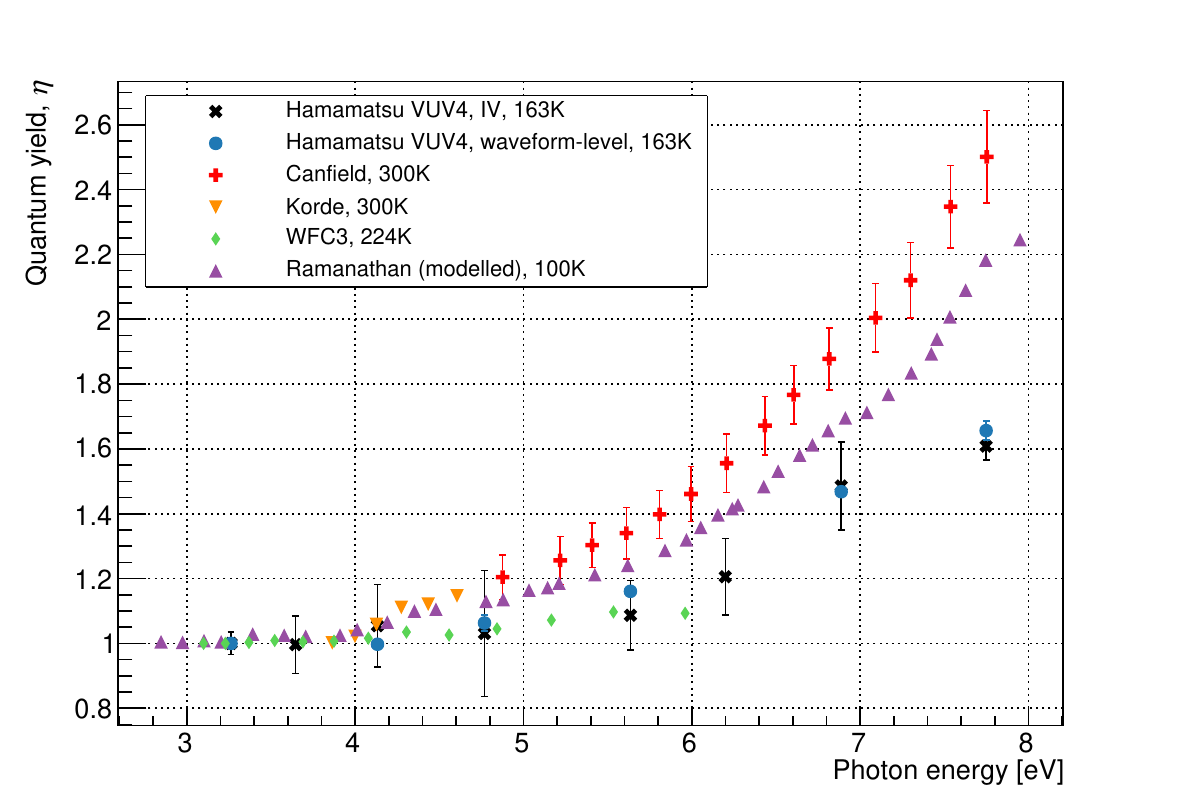}
    \caption{Values for $\eta$ as a function of photon energy, calculated using waveform-level and IV measurements.  Previously reported values from Canfield \textit{et al.} \cite{canfield1998absolute}, Korde \textit{et al.} \cite{korde_quantum_1987}, and Borders \textit{et al.} \cite{Borders_wfc3_2010} are given for comparison, as well as modelled values from Ramanathan \textit{et al.} \cite{ramanathan_ionization_2020}. }
    \label{fig:litcomparison}
\end{figure}
Our measured values of $\eta$ are shown in figure \ref{fig:litcomparison} with previously measured \cite{canfield1998absolute,korde_quantum_1987,Borders_wfc3_2010} and modelled \cite{ramanathan_ionization_2020} data. Our data agree well with those of Borders \textit{et al.} \cite{Borders_wfc3_2010} but are significantly lower than those given by Canfield \textit{et al.}  \cite{canfield1998absolute}. These new data suggest that the pair creation energies for silicon are higher for depositions with energies below the `UV-gap' than for photon energies in the x-ray range or above. This is contrary to previous measurements, which suggested that lower-energy depositions resulted in lower pair creation energies. The agreement between the measured values produced by the different methods detailed here gives confidence in their accuracy, and agreement between the two devices measured here and the CCDs measured in \cite{Borders_wfc3_2010} indicates that the quantum yield does not vary significantly between different devices. As the quantitative values in \cite{ramanathan_ionization_2020} were produced by fitting to the previously measured data of Canfield \textit{et al.} \cite{canfield1998absolute}, this work may necessitate a correction to the modelled pair creation probabilities at recoil energies below 10eV. The higher energies required to produce more than a single electron-hole pair will have implications for the sensitivity of experiments with thresholds of $2e^-$ or higher, and is expected to impact the physics reach that can be achieved using such measurements. This data should also be considered when assessing the performance of novel silicon photodetectors with quantum efficiencies close to or above unity.

\backmatter

\bmhead{Acknowledgements}

The authors gratefully acknowledge the support of Nicolas Massacret and Lars Martin in the design and operation of the experimental setup. The authors acknowledge support from Canadian Foundation for Innovation Fund (CFI) 2017. Additional support was provided by a grant from the Natural Sciences and Engineering Research Council of Canada for the nEXO project. 

\bibliography{citation}

\end{document}